\documentclass[aps,physrev,twocolumn,groupedaddress]{revtex4-2}

\usepackage{siunitx}
\usepackage{graphicx}
\usepackage{amsmath}
\usepackage{amssymb}
\usepackage{chemformula}
\usepackage{hyperref}

\begin{document}

% Use the \preprint command to place your local institutional report
% number in the upper righthand corner of the title page in preprint mode.
% Multiple \preprint commands are allowed.
% Use the 'preprintnumbers' class option to override journal defaults
% to display numbers if necessary
%\preprint{}

\title{Signatures of Dynes density of states in the terahertz response of atomic layer deposition grown superconducting NbN thin films}

\author{Frederik Bolle$^{1}$, Yayi Lin$^1$, Ozan Saritas$^{1}$, Martin Dressel$^{1,2}$, Ciprian Padurariu$^{3}$,\\ Sahitya Varma Vegesna$^{4,5}$, Nitesh Yerra$^{4}$, Heidemarie Krüger$^{4,5}$,  and Marc Scheffler$^{1,2}$ }

\affiliation{$^1$1. Physikalisches Institut, Universität Stuttgart, 70569 Stuttgart, Germany}
\affiliation{$^2$ Center for Integrated Quantum Science and Technology (IQST),  Universität Stuttgart, 70569 Stuttgart, Germany}
\affiliation{$^3$Institute for Complex Quantum Systems and IQST, Universität Ulm, 86069 Ulm, Germany}
\affiliation{$^4$Leibniz Institute of Photonic Technology, 07745 Jena, Germany}
\affiliation{$^5$Institute for Solid State Physics, Friedrich-Schiller Universität Jena, 07743 Jena, Germany}

\date{\today}
\begin{abstract}
The frequency-dependent complex optical conductivity $\hat{\sigma}(f)$ reflects key properties of superconductors, such as the energy gap $2\Delta$ in the density of states (DOS) and the superfluid density $n_\mathrm{s}$. 
For disordered superconductors,  $\hat{\sigma}(f)$ often can be described within Bardeen-Cooper-Schrieffer (BCS) theory, while in corresponding tunneling experiments, deviations in the observed DOS typically require modelling by the phenomenological Dynes formula. 
The implications of such Dynes DOS for optics were rarely discussed so far.
Here we probe the tera\-hertz $\hat{\sigma}(f)$ of superconducting NbN thin films with thicknesses ranging from $\SIrange{4.5}{20}{\nano\meter}$, which were grown by atomic layer deposition (ALD). Our frequency range from \SIrange{0.3}{2.1}{\tera\hertz} covers energies below and above $2\Delta$.
For \SI{20}{\nano\meter} thick NbN, we find  in $\hat{\sigma}(f)$ distinct deviations from the BCS model, including a step-like characteristic in the absorption at $\Delta_0$, i.e.\ half the zero-temperature spectral gap. 
These observations can be fully captured by introducing a small and temperature-independent Dynes pair-breaking rate $\Gamma \sim 0.036\,\Delta_0$. 
For the other films, we also observe signs of Dynes-type DOS, and we discuss the evolution of $2\Delta_0$, $n_\mathrm{s, 0}$, and $\Gamma$ as function of film thickness.

\end{abstract}

\maketitle

\section{Introduction}

When a conventional s-wave superconductor is cooled below its critical temperature $T_\mathrm{c}$, a gap of size $2\Delta$ opens up in the density of states (DOS) due to the formation of Cooper pairs. 
According to the Bardeen-Cooper-Schrieffer (BCS) theory, Cooper pairs can be broken by thermal excitation or radiation with sufficient energy $E > 2\Delta$. 
Therefore, at very low temperatures $T \ll T_c$, no absorption channels exist within the superconducting gap $2\Delta_0 = 2\Delta(T=0)$, since the DOS is exactly zero.

Experimentally, the energy-dependent density of states $ N(E)$ of a superconductor can be probed by means of tunneling spectroscopy, either using scanning tunneling microscopy in spectroscopy mode or using junctions with thin dielectric films as tunneling barriers. 
In the framework of disordered and ultra-thin superconductors, such measurements typically reveal a non-zero DOS at zero bias even at temperatures far below the critical temperature \cite{Chockalingam2009, Szabo2016, Proslier2008} and in zero magnetic field, which is not accounted for within the BCS model. 
To quantitatively describe the emergence of these in-gap states, one can employ the Dynes model \cite{Dynes1978}, which introduces a phenomenological scattering rate $\Gamma$ that quantifies additional pair-breaking mechanisms in superconductors
\begin{align}
    N(E) = N_0 \operatorname{Re}\left(\frac{E + i\Gamma}{\sqrt{\left(E + i\Gamma \right)^2 - {\Delta}^2}} \right) . \label{eq:dynes-density-of-states}
\end{align}
Here $N_0$ is the normal-state DOS at the Fermi level, and $\Delta$ is the ideal superconducting gap within the BCS model. %, and $\omega = 2 \pi f$ is the angular frequency. 
These modifications to the single-particle DOS have direct consequences for the dynamical conductivity \cite{Herman2017}, including a characteristic step-like absorption feature at half of the spectral gap $\Delta$. Despite the wide use of the Dynes model in describing tunneling data \cite{Szabo2016, Chockalingam2009, Boschker2020, Haskova2018}, a direct observation of this postulated absorption step by optical means has not been reported yet.  

The superconducting electrodynamics of sputtered or pulsed laser deposited NbN thin films have widely been investigated \cite{Karecki1983, Kornelsen1991, Ikebe2009, Sindler2010, Beck2011, Matsunaga2012, Pracht2013, Sherman2015, Cheng2016, Sindler2022}.
In this work, we study a series of atomic layer deposited NbN thin films \cite{Linzen2017, Deyu2025} using tera\-hertz (THz) spectroscopy to directly probe the complex conductivity $\hat{\sigma}(f)$ over a broad range of frequencies $f$ and temperatures $T$.

\section{Experiment}

NbN films of thicknesses $d$ ranging from \SIrange{4.5}{20}{\nano\meter} were deposited on \SI{530}{\micro\meter} thick $10\, \times\, 10\,$mm$^2$ R-plane sapphire (\ch{Al2O3}) using atomic layer deposition (ALD). For more details regarding the precise deposition parameters and used precursors, consider the report by Linzen \textit{et al}.\ \cite{Linzen2017}. Various sample properties are listed in \autoref{tab:samples}.
As a first characterization of the thin films, transport measurements were conducted in van der Pauw geometry to obtain the transition temperature $T_c$ and normal-state resistivity $\rho_0$. 
To study the optical response in the THz regime \cite{Pracht2013}, a commercial THz time-domain spectrometer (TeraPulse 4000) as well as a frequency-domain spectrometer were used in transmission geometry. In the latter, frequency-tunable backward wave oscillators (BWO) generate monochromatic, coherent light in the THz regime, which is transmitted through the sample and detected using a Golay cell or a \ch{^4He} cooled bolometer \cite{Pracht2013}. This approach has been used extensively to study the low-energy optical conductivity of superconductors \cite{Pracht2016, Simmendinger2016, Kang2011}. The photon energy $E = h f$, with Planck's constant $h$, is the relevant energy scale of the optical experiments below. 
For convenience and closer connection to the THz experiments with $f$ as common tuning parameter, we will relate energy scales such as $\Delta$ to either $E$ or $f$, depending on the context.

\section{Results}

\subsection{Transport properties and optical conductivity}

\begin{table*}
    \centering
        \caption{Summary of characteristic material properties of superconducting NbN samples such as the thickness $d$, the sheet resistance $R_\mathrm{s,DC}$, the critical temperatures obtained from DC $T_{\mathrm{c}, \text{DC}}$ and THz measurements $T_{\mathrm{c}, \text{THz}}$, the spectral gap $2\Delta_0$, the coupling ratio $2\Delta_0/ (k_\mathrm{B} T_\mathrm{c,DC})$, the superfluid density  $n_\mathrm{s,0}$, the London penetration depth $\lambda_\mathrm{L,0}$, the sheet kinetic inductance $L^{\square}_\mathrm{kin,0}$, and the pair-breaking rate $\Gamma/\Delta_0$. $T_\mathrm{c, DC}$ and $R_\mathrm{s, DC}$ have been obtained through DC transport measurements.  For samples with  $d < \SI{7.5}{\nano\meter}$, $\Delta_0$ has been solely determined from frequency-domain measurements.}
    \begin{tabular}{c|c|c|c|c|c|c|c|c|c|c|c} \hline
         $d$ & $R_\mathrm{s,DC}$  & $\rho_\mathrm{dc}$ & $T_\mathrm{c,DC}$ & $T_\mathrm{c,THz}$ & $2\Delta_0$  & $2\Delta_0$  &  $2\Delta_0/ (k_\mathrm{B} T_\mathrm{c,DC})$ & $n_\mathrm{s,0} \cdot 10^{25}$  & $\lambda_\mathrm{L,0}$ & $L^{\square}_\mathrm{kin,0}$  & $\Gamma/\Delta_0$ \\ 
         \textrm{[nm]} &  [$\Omega$/sq]& [$\mu\Omega$m] &  [K]&  [K] &  [THz] & [meV] &  &  $[\text{m}^{-3}]$ &  \textrm{[nm]} &  [pH/sq] &  \\ \hline
         4.5 & 1445 & 6.50 & 8.49&8.63&0.605&2.50&3.42&3.04&848&259&0.028\\ \hline
         5 & 695 & 5.55 & 9.74&9.74&0.707&2.92&3.49&5.08&740&139&0.019\\ \hline
         7.5 & 413 & 3.10 & 12.14&11.63&0.971&4.02&3.84&10.85&532&43.6&0.017\\ \hline
         10 & 256 & 2.55 & 12.89&12.72&1.073&4.44&4.00&14.55&493&24.4&0.010\\ \hline 
         20 & 112& 2.23 & 13.34&13.60&1.124&4.65&4.05&16.87&443&10.5&0.036
    \end{tabular}
%} 
\label{tab:samples}
\end{table*}

\begin{figure}
\includegraphics[width=1\linewidth]{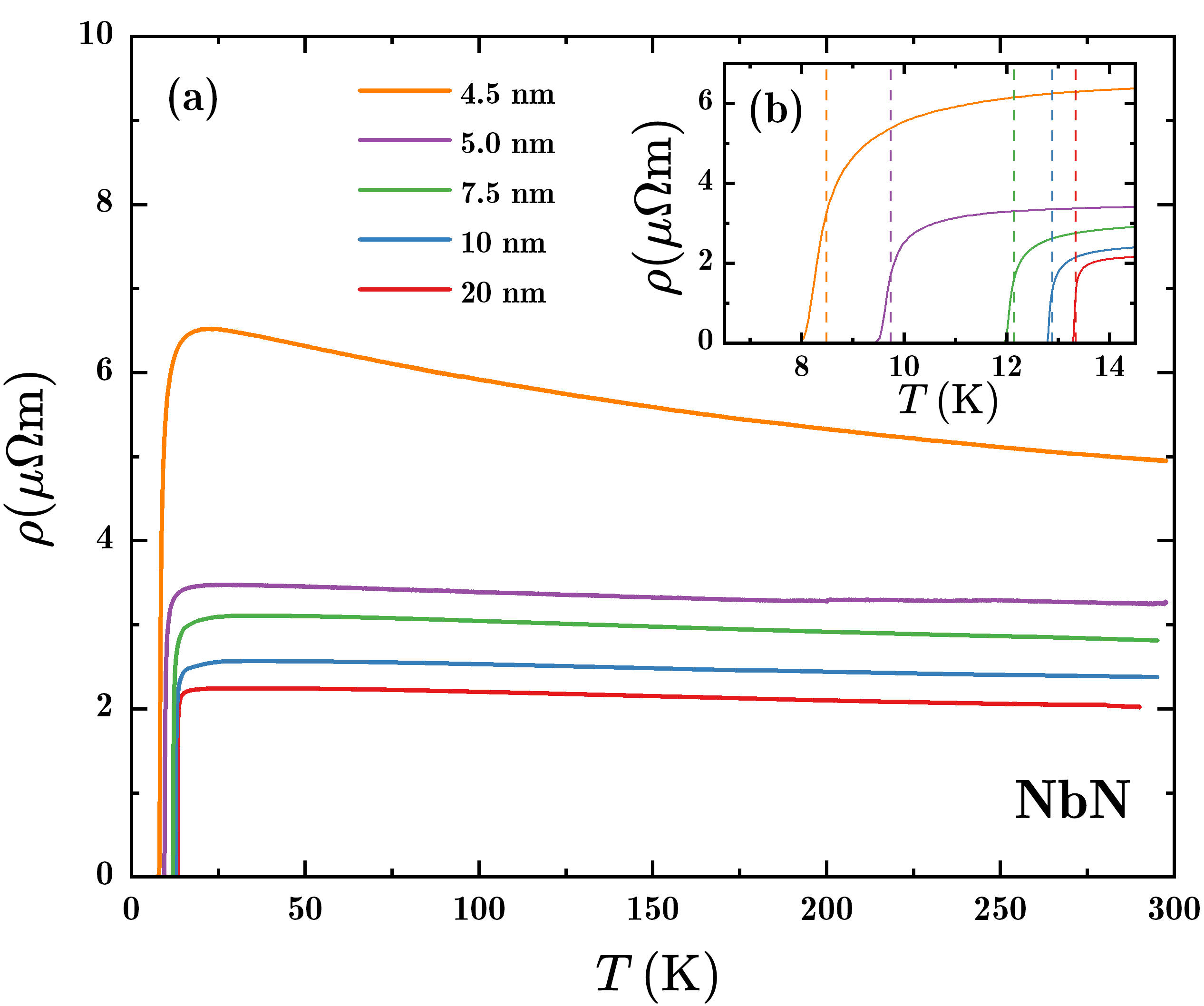}
\caption{The main panel (a) shows the temperature-dependent resistivity $\rho(T)$ for all studied films, while the inset (b) displays a magnified view of the behavior near the superconducting transition temperature $T_\mathrm{c}$, where $T_c$ is defined via the 50\% criterion with respect to the maximum of the normal-state resistivity, 
as indicated by dashed lines. 
\label{fig:DC}}
\end{figure}

Results of transport measurements, displayed in \autoref{fig:DC}, show that all five films become superconducting with critical temperatures ranging from $\SIrange{8.5}{13.3}{\kelvin}$. 
With decreasing film thickness, the normal-state resistivity increases, and the critical temperature is suppressed significantly \cite{Semenov2009, Kamlapure2010, Ivry2014, Bouteiller2025}. Additionally the width of the transition increases as the film dimensions become increasingly two-dimensional.

\begin{figure}
\centering
\includegraphics[width=1\linewidth]{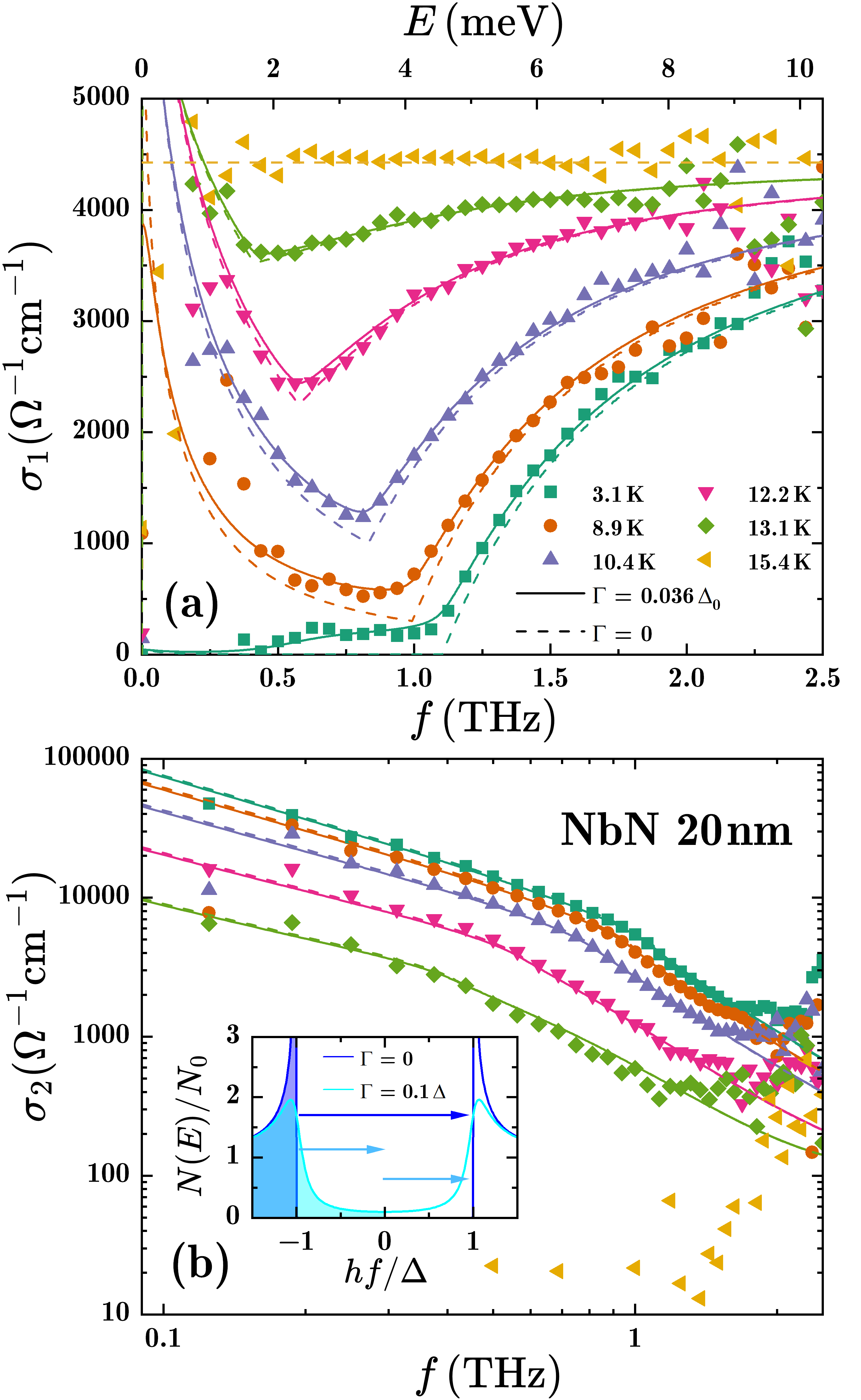}
\caption{
Complex optical conductivity $\hat{\sigma} = \sigma_1 + i \sigma_2$ of a $\SI{20}{\nano\meter}$ NbN thin film with $T_c \approx \SI{13.3}{\kelvin}$, obtained with THz time-domain spectroscopy (TDS), where panel (a) shows the real part $\sigma_1$ on linear scales and (b) the imaginary part $\sigma_2$ on logarithmic scales. 
Solid lines represent fits according to the Dynes model \cite{Herman2017} with a finite and constant pair-breaking rate of $\Gamma = 0.036 \, \Delta_0$, while dashed lines are calculated according to the Mattis-Bardeen model \cite{Zimmermann1991}, corresponding to $\Gamma = 0$. The strongest deviations between the Dynes and Mattis-Bardeen models occur near the temperature-dependent minimum of $\sigma_1(f)$. The normal-state conductivity at \SI{15.4}{\kelvin} is sketched as a dashed, horizontal line in (a) and corresponds to the value $\sigma_\textrm{dc} =  4460\ \Omega^{-1}\mathrm{cm}^{-1}$ obtained from transport measurements.
The inset in (b) shows the superconducting density of states near the Fermi energy in the limit $T=0$. In the BCS case ($\Gamma =0$), only excitations with $hf > 2 \Delta$ are allowed (dark blue arrow). In the Dynes case (here: $\Gamma = 0.1 \Delta$), also excitations at lower energies are possible, and the transitions indicated by light blue arrows, with $hf = \Delta$, lead to the step near $\Delta$ in $\sigma_1(f)$.
\label{fig:sigma-dynes}}
\end{figure}

The complex optical conductivity of the \SI{20}{\nano\meter} NbN film obtained using THz time-domain spectroscopy (TDS) is shown in \autoref{fig:sigma-dynes}. Above the critical temperature of around $\SI{13.3}{\kelvin}$, the film shows a frequency-independent real part $\sigma_1$ of the complex optical conductivity $\hat{\sigma} = \sigma_1 + \mathrm{i} \sigma_2$, indicating a normal-state Drude-type scattering rate $\Gamma_n$ far outside our accessible spectral range, as expected for a dirty superconductor above $T_c$. Below the critical temperature, the low-frequency $\sigma_1$ is significantly suppressed due to the pairing of electrons into Cooper pairs. The corresponding minimum in $\sigma_1$ is readily identified as the spectral gap $2\Delta$ above which Cooper pairs are broken up by incoming THz photons into individual quasiparticles. Simultaneously, a clear $1/f$ dependence develops in the imaginary part $\sigma_2$ for frequencies below the spectral gap, which increases in magnitude for lower temperatures \cite{Dressel2002}. 
To describe the dynamical conductivity, we first use the model by Zimmermann \textit{et al}.\ \cite{Zimmermann1991}, which is a description of the Mattis-Bardeen equations generalized for samples of arbitrary purity. Motivated by the frequency-independent normal-state conductivity, we set the impurity parameter
$y = \hbar \Gamma_n /(2\Delta) = 500$ (impure limit) \cite{Zimmermann1991}, with reduced Planck's constant $\hbar$.
The model fits our data reasonably well and allows us to extract superconducting properties, such as the temperature-dependent spectral gap $2\Delta(T)$ and the superfluid density $n_s(T)$, which will be discussed below. All obtained parameters from transport and optical measurements are summarized in \autoref{tab:samples}.

\begin{figure}
\centering
\includegraphics[width=1\linewidth]{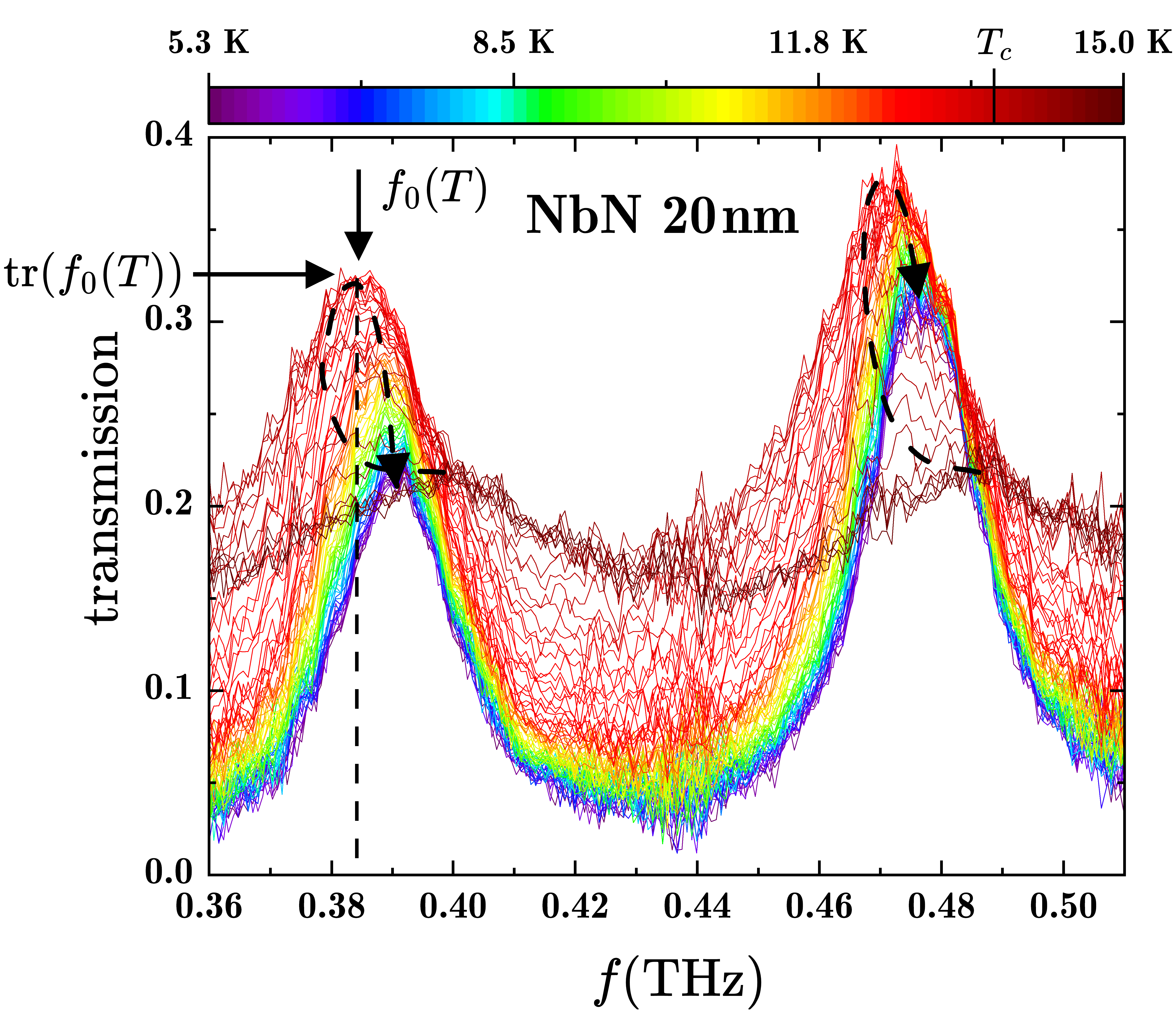}
\caption{Fabry-Pérot (FP) resonances, observed for numerous temperatures in THz frequency-domain spectroscopy (FDS) on \SI{20}{\nano\meter} NbN for two modes, with FP frequencies $f_{0,n}=\SIlist{0.4;0.49}{\tera\hertz}$ in the metallic state at $T = \SI{15}{\kelvin}$. Arrows indicate the positions of the temperature-dependent resonance frequency and its corresponding transmission value. Additionally, dashed arrows are guides to the eye to indicate the shift of the modes as the temperature is lowered.}\label{fig:raw-FDS}

\end{figure}

\subsection{Signatures of Dynes-type density of state}

\begin{figure}
    \centering
    \includegraphics[width=1\linewidth]{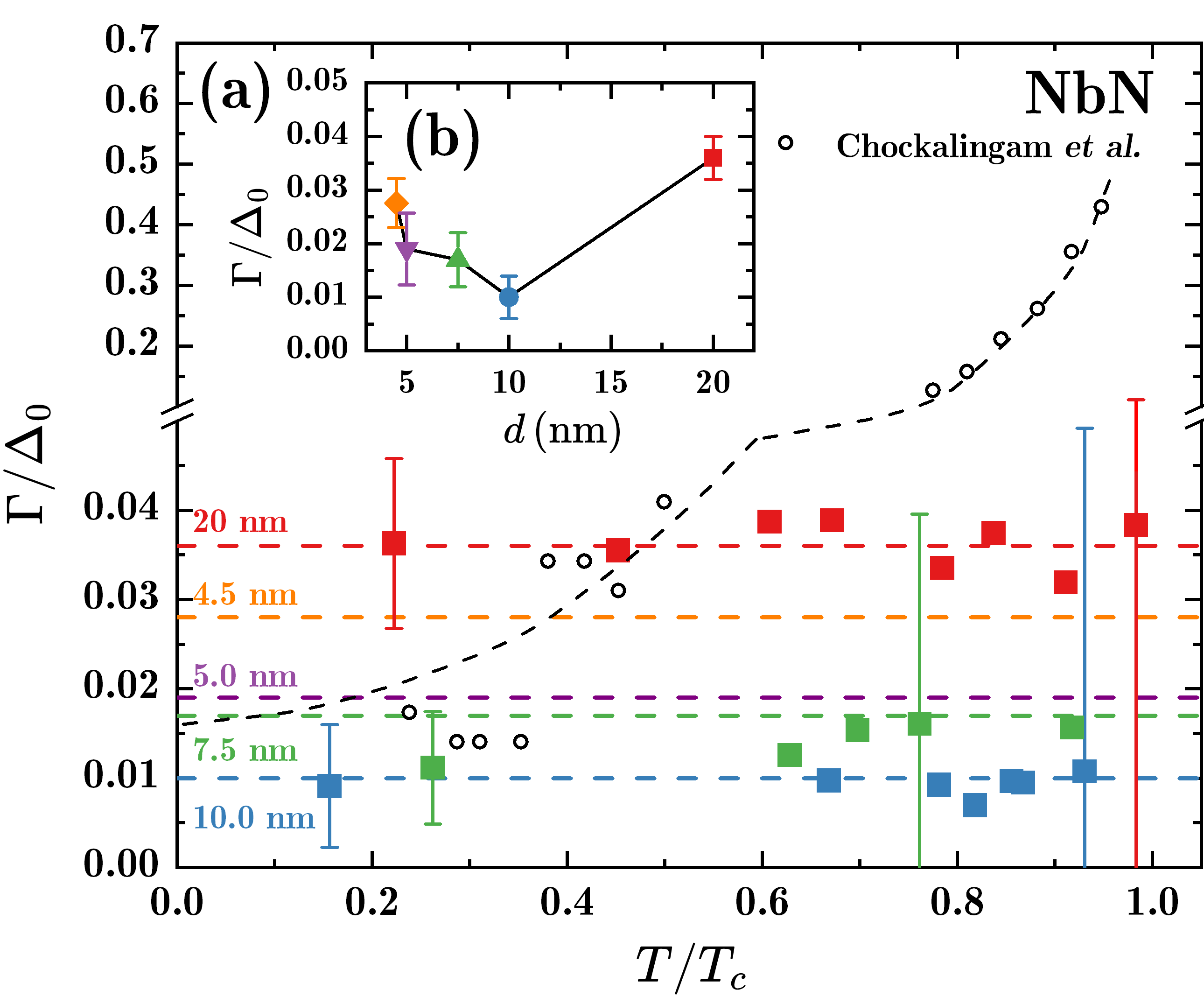}
    \caption{(a) Pair-breaking rates $\Gamma$ for five NbN samples normalized to their zero-temperature energy gap $\Delta_0$, from both TDS (discrete temperature points) and FDS (dashed lines) studies. For comparison, pair-breaking rates obtained by Chockalingam \textit{et al.\ }\cite{Chockalingam2009} from tunneling spectroscopy measurements on NbN are shown as open diamonds with a dashed line as a guide to the eye. The inset (b) displays $\Gamma/\Delta_0$, as determined from FDS, for the different film thicknesses. Error bars are shown for all FDS results in (b) and for exemplary TDS results in (a). A detailed error discussion can be found in the Supplemental Material \cite{Supplemental}.
    \label{fig:gamma}}
\end{figure}

Considering the $\sigma_1$ spectrum at lowest temperature $T = \SI{3.1}{\kelvin}$ in \autoref{fig:sigma-dynes}, we find that the Zimmermann fit does not fully describe the data. We observe an onset of absorption starting at half the spectral gap $2\Delta$, which is not captured by the standard BCS model for the dynamical conductivity. 
To describe the additional losses seen in the experiment, we use the model for the optical conductivity of superconductors with pair-breaking processes as described by Herman and Hlubina \cite{Herman2017} and based on the Dynes DOS, \autoref{eq:dynes-density-of-states}, as illustrated in the inset of \autoref{fig:sigma-dynes}(b).
This modification leads to a smearing of the energy gap together with the emergence of a distinct absorption step at $\Delta$, which can be understood when considering the modified DOS. Due to the gapless nature of such superconductors, absorption is possible at arbitrarily low frequencies similar to a normal metal. 
The joint density of states and correspondingly $\sigma_1(f)$  increase at both steps: one at $\Delta$ (corresponding to excitations from the DOS maximum around $-\Delta$ to the Fermi energy $E_\textrm{F}$ at $E_\textrm{F} = 0$ or from $E_\textrm{F}$ to the DOS maximum around $+\Delta$) and the other at $2\Delta$ (corresponding to excitations between the two DOS maxima at $-\Delta$ and $+\Delta$).
We model our conductivity spectra with a two-parametric fit for all temperatures to obtain the temperature-dependent energy gap $\Delta(T)$ and the pair-breaking rate $\Gamma(T)$, which is basically constant $\Gamma = 0.036\,\Delta_0$, as shown in \autoref{fig:gamma}. 
By using the Dynes model we can achieve a very good description of the dynamical conductivity in \autoref{fig:sigma-dynes}, capturing the step-like feature observed at $\SI{3.1}{\kelvin}$, as well as the systematic deviation at higher temperatures.

\begin{figure}
    \centering
    \includegraphics[width=1\linewidth]{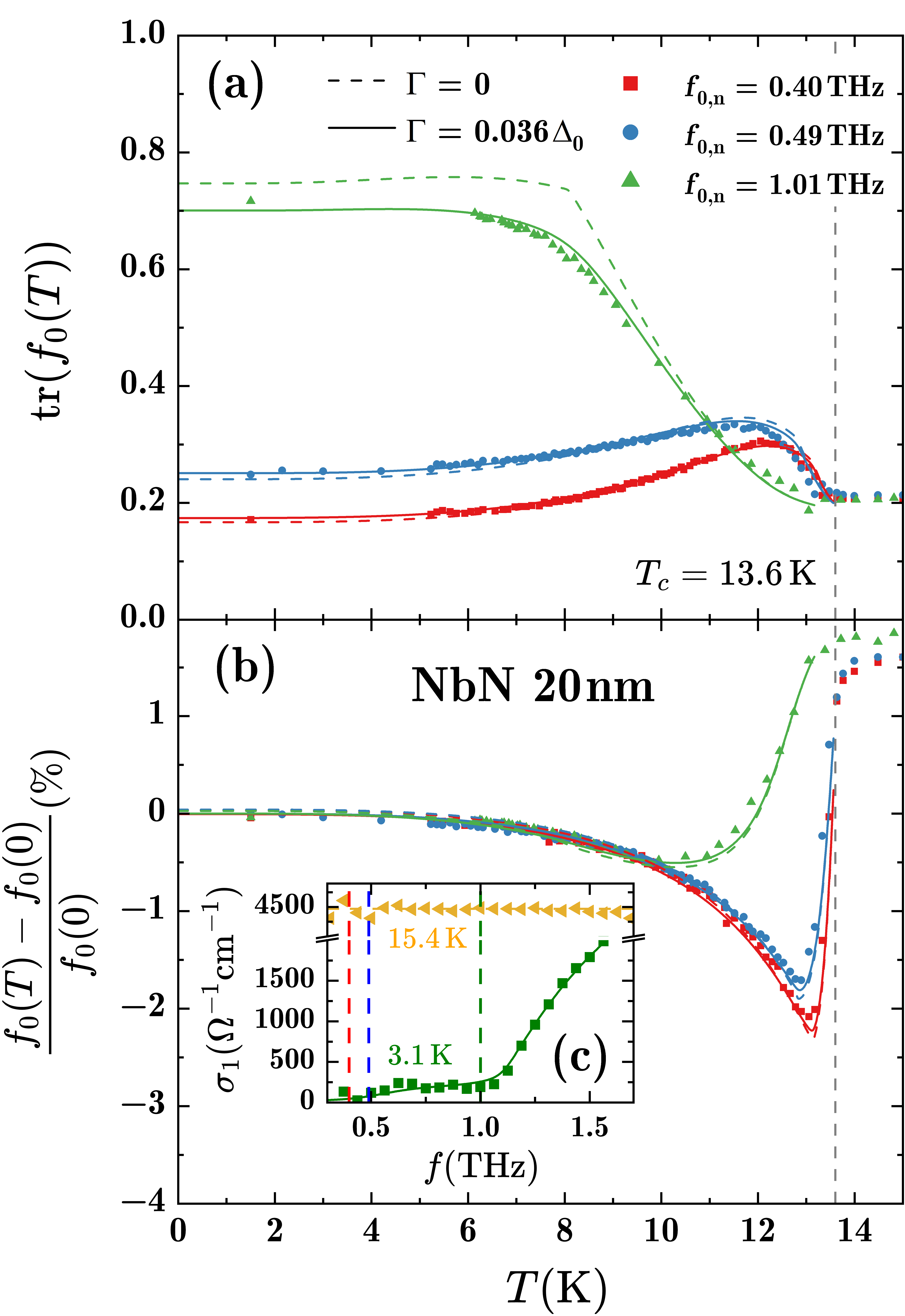}
    \caption{Frequency-domain spectroscopy (FDS) studies of the temperature-dependent transmission $\text{tr}(f_0(T))$ at temperature-dependent resonance frequency $f_0(T)$ and fractional resonance frequency shift $\frac{f_0(T) - f_0(0)}{f_0(0)}$ displayed in panel (a) and (b) respectively. Solid lines represent fits according to the Dynes model \cite{Herman2017} with a finite pair-breaking rate of $\Gamma = 0.036 \, \Delta_0$, while dashed lines are modeled according to the Mattis-Bardeen model \cite{Zimmermann1991}, corresponding to $\Gamma = 0$. The vertical line at $T = \SI{13.6}{\kelvin}$ indicates $T_c$ obtained from the THz measurements. The inset (c) shows the real part of the optical conductivity $\sigma_1$, obtained from TDS, at $\SI{3.1}{\kelvin}$ and in its normal state at $\SI{15.4}{\kelvin}$, with dashed vertical lines indicating the frequency positions of the Fabry-Pérot modes studied by FDS.\\
}
    \label{fig:nu-tr-analysis}
\end{figure}

To further support this observation of Dynes-type features in the superconducting THz electrodynamics of our films, we turn to frequency-domain spectroscopy (FDS). 
As an example the raw transmission spectrum of the \SI{20}{\nano\meter} NbN film for numerous temperatures is shown in \autoref{fig:raw-FDS}. 
We observe the periodic Fabry-Pérot (FP) transmission pattern, caused by constructive and destructive interference in the dielectric substrate and modified by the temperature-dependent superconducting properties of the thin film.
As the film is cooled below $T_\mathrm{c}$, strong changes can be observed in absolute transmission and in resonance frequency. 
We treat each FP mode as a resonating mode, with a characteristic resonance frequency $f_0$ and an absolute transmission at the maximum $\text{tr}(f_0)$.

Here the THz photon energy is comparable to $\Delta_0$, and the in-gap absorption seen previously in TDS should also appear in this FDS experiment. 
The result of this investigation for three FP modes is shown in \autoref{fig:nu-tr-analysis}, with the absolute transmission in panel (a) and the normalized change in resonance frequency in panel (b). 
We model the FP transmission pattern using an analytic expression for the transmission coefficient \cite{Dressel2002} combined with the Dynes model for optical conductivity. 
Motivated by the temperature-independent pair-breaking rate seen by TDS, we assume a constant $\Gamma(T) = \Gamma$. 
Other characteristic parameters, such as the spectral gap and normal-state conductivity correspond to the ones obtained from TDS. 
As evident from \autoref{fig:nu-tr-analysis}, we observe a very convincing match between theory and experiment, fully reproducing the temperature and frequency dependence of the modes. 
As the film is cooled below $T_c$, a gap opens up leading to a suppression of $\sigma_1(f)$. 
As a consequence, for modes close to the spectral gap $2\Delta_0$, in our case for $f = \SI{1}{\tera\hertz}$, where the transmission is dominated by changes in $\sigma_1$, the transmission strongly increases upon cooling. 
For lower frequency modes, an interplay between $\sigma_1$ and $\sigma_2$ leads to the emergence of a distinct maximum of the transmission just below the critical temperature (not to be confused with the coherence peak in $\sigma_1(T)$ at low frequencies \cite{Steinberg2008}). 
This maximum is at much lower temperatures and much weaker for the mode at $\SI{1}{\tera\hertz}$, where additionally a clear change of slope in the absolute transmission (best seen in the Mattis-Bardeen prediction at around $T = \SI{8}{\kelvin}$) indicates the crossing of the FP mode with the kink in $\sigma_1(f)$ at the spectral gap $2\Delta(T)$.
For the highest frequency mode at around $\SI{1}{\tera\hertz}$ ($E \approx 1.8 \Delta_0$), the absolute transmission of the mode clearly shows strong deviations from the BCS prediction, indicating the existence of an additional pair-breaking mechanism, which leads to the excess sub-gap conductivity. 
The magnitude of the pair-breaking rate determined from this FP investigation in FDS, $\Gamma = 0.036\,\Delta_0$, matches TDS on the same film and confirms the temperature-independence of the pair-breaking rate. 
THz data obtained on the other four samples as well as a discussion of experimental error sources and their effect on the analysis are presented in the Supplemental Material \cite{Supplemental} (see also references \cite{Neu2018, Glover1957, Pronin1998, Bachar2015, Gorshunov2007} therein).

Combining both TDS and FDS studies, we plot in \autoref{fig:gamma} the obtained pair-breaking rates $\Gamma$ for the five NbN samples, normalized to $\Delta_0$. For all of them we find temperature-independent $\Gamma$. This is in contrast to tunneling spectroscopy results on a NbN film with comparable level of disorder previously reported by Chockalingam \textit{et al}.\ \cite{Chockalingam2009}, which we have included in \autoref{fig:gamma} as comparison. 
For low temperatures the magnitude of the pair-breaking rate from THz and tunneling measurements are similar, but strong deviations emerge for higher temperatures, where tunneling measures values of the pair-breaking rate of up to 50\% of $\Delta_0$, while the values from our optical measurements remain small and temperature-independent. 
One relevant parameter here is the size of the volume that is probed, as local changes in resistivity can lead to an inhomogeneous energy landscape of the superconductor \cite{Carbillet2020}. 
In contrast to scanning tunneling spectroscopy with sub-nm spatial resolution, both our THz study and the tunneling experiment in \cite{Chockalingam2009} average over comparable lateral dimensions defined by the THz spot size (comparable to wavelength) and the junction area, respectively.
 One difference, though, is the spectroscopy probing depth: while THz transmission probes the complete film thickness and thus is mostly sensitive to the bulk of the NbN thin film, tunneling is inherently surface sensitive and in the case of thin-film junctions might also be affected by details of the barrier.
Another, generally relevant possible cause for observed differences in the Dynes scattering parameter are differences in thin-film growth, which affect the microscopic film structure.

\subsection{Superconducting properties}

\begin{figure}
\includegraphics[width=1\linewidth]{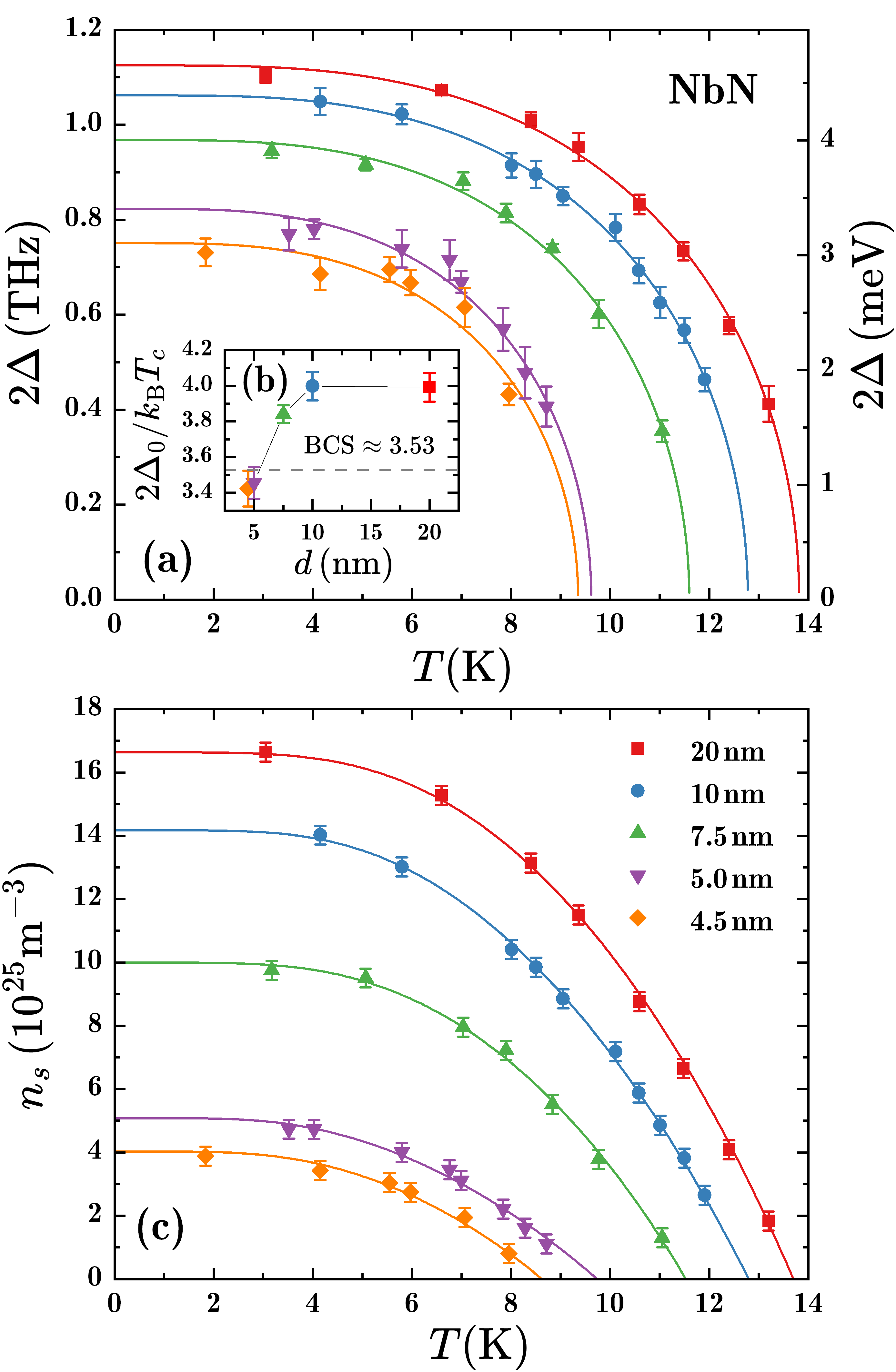}
\caption{(a) Temperature-dependent superconducting energy gap $2\Delta(T)$ and (c) superfluid density $n_s(T)$ obtained from using the Dynes model. The temperature dependence of the energy gap is modeled using a BCS approximation $\Delta(T) = \Delta(0) \tanh({1.74\sqrt{T_c/T - 1}})$ while the superfluid density is modeled using the procedure described in \cite{Dutta2022}. The inset in (b) depicts the gap ratio $\frac{2\Delta_0}{k_\mathrm{B} T_c}$ with the dotted line representing the weak coupling BCS limit. \label{fig:sc-param}}
\end{figure} 

Lastly we discuss the superconducting properties, in particular the energy gap and superfluid density shown in \autoref{fig:sc-param}(a) and (c), respectively, for the five NbN samples. To determine the density of superconducting charge carriers, we use the Kramers-Kronig relations to relate the magnitude of the $\delta$-distribution in $\sigma_1(f = 0)$ to the imaginary part $\sigma_2(f)$
\begin{align}
    n_s = \frac{2\pi m^*}{e^2} \lim_{f \rightarrow 0} f \sigma_2(f),
\end{align}
where $m^* = m_e$ is the effective mass of quasiparticles in NbN. The temperature dependence of the energy gap is well described within the BCS model. For thicker films, the gap ratio $2\Delta_0/(k_B T_c)$ of our NbN films is significantly higher ($\approx 4.0$) than the weak coupling limit of $3.528$, indicating strong electron-phonon coupling \cite{Kornelsen1991}. Reducing the film thickness suppresses its superconducting properties due to increased electron-electron repulsion which suppresses the phonon-mediated attractive pairing. 
To classify our results for the coupling ratio, we consider the abundant literature on NbN, with works reporting conflicting behaviors as $T_\mathrm{c}$ is suppressed with increasing disorder, typically in conjunction with reduced film thickness \cite{Chockalingam2009, Kang2011, Henrich2012, Cheng2016}. 
In particular the coupling ratio, which is directly related to the magnitude of the energy gap and its critical temperature, can show varying behaviors as disorder is increased. A report by Chand \textit{et al}.\ reveals the complex nature of the phase diagram of strongly disordered NbN thin films \cite{Chand2012}. 
As one increases the disorder of the system, first superconductivity is suppressed due to gradual reduction of pairing interactions after which an intermediate disorder regime follows, which is governed by the suppression of the superfluid stiffness $J$. This intermediate regime is concomitant with the emergence of a pseudo-gap state surviving even above the critical temperature, which persists until (maybe even beyond) the destruction of superconductivity at $k_f l \approx 1$, with Fermi wavevector $k_f$ and mean free path $l$. 
Beyond critical levels of disorder, the insulating regime of NbN is characterized by incoherent superconducting islands, reminiscent of the situation in other disordered superconductors such as \ch{TiN} \cite{Kamlapure2013}, \ch{InO_x} \cite{Sacepe2011} and granular aluminum \cite{Pracht2016}. 
The rich phase diagram of NbN indicates that multiple processes of both fermionic (electronic interactions) and bosonic nature (phase coherence between Cooper pairs) play a role when approaching the transition to its disorder-induced insulating state. 
From the continuous reduction of the coupling ratio and the absence of phenomena indicating a pseudo-gap, we conclude that the suppression of superconductivity within the disorder regime of the films studied in this work is mainly governed by electronic interactions, e.g. they lie in the fermionic regime of the superconductor-insulator transition of NbN.

\section{Discussion}

We find that for the NbN films under study, conventional BCS electrodynamics cannot describe the THz data whereas inclusion of a finite Dynes pair-breaking parameter $\Gamma$ leads to excellent agreement.
While the Dynes formula has been useful as a phenomenological model that yields an in-gap density of states, the microscopic origin of the Dynes parameter remains unclear. A number of competing phenomena have been proposed to contribute to the Dynes parameter. 
Two main mechanisms are valid in most conventional superconducting materials: (i) inelastic electron scattering that leads to Cooper-pair breaking \cite{Kaplan1976, Mikhailovsky1991} and (ii) magnetic scattering \cite{Herman2016,Herman2017,Herman2018} that lowers the local single-particle excitation energy. 
Other mechanisms have also been identified in materials with spatially inhomogeneous or exotic order parameters, or in materials with different superconducting order parameters in different bands \cite{Boschker2020}. 
Inelastic electron scattering can occur either due to electron-phonon scattering, or due to electron-electron scattering.
Since inelastic electron scattering is also responsible for the normal-state conductance of a metal, the resulting contribution to the Dynes parameter is expected to have a similar magnitude and temperature dependence when extrapolated above the critical temperature. 
From the temperature-independence of the pair-breaking rate seen by our two independent THz measurement procedures, we infer that inelastic electron scattering processes are not the prevailing explanation for the emergence of in-gap states in NbN.
Magnetic scattering at a localized impurity induces a single, discrete, and well-localized Yu-Shiba-Rusinov state inside the superconducting gap, and not a continuum of in-gap states. 
However, it has been shown that hybridization of such states due to a random density of magnetic scattering centers can lead to an in-gap density of states consistent with the Dynes formula \cite{Herman2016, Herman2017, Herman2018}. Additionally, a recent theoretical work by Yang and Wu \cite{Yang2024} predicts an interband transition between Shiba impurity bands across the energy gap, which leads to an optical absorption feature close to $\Delta$.  
In tunneling studies on ultra-thin superconducting films it was shown that the Dynes parameter increases significantly when the film thickness is reduced \cite{Noat2013, Szabo2016, Chockalingam2009, Tamir2022}. This property of ultra-thin films was measured both for NbN films \cite{Noat2013, Kamlapure2010} and for MoC films \cite{Haskova2018}. The proposed explanation is that the interface of the superconducting thin films presents an excess of magnetic impurities, such that for thinner films, their effect on the Dynes in-gap density of states is more pronounced.
From our measurements, we conclude that the pair-breaking rate remains small in its magnitude (only few percent of the energy gap) and does not show a systematic dependence on film thickness. Although we do observe a slight increase (excluding the \SI{20}{\nano\meter} sample) in the pair-breaking rate when reducing the film thickness, as seen in the inset \autoref{fig:gamma}(b), the effect is too weak to make a conclusive statement, leaving the question open where the pair-breaking fields predominantly reside within the superconducting film.

Our interpretation centers around the Dynes approach and how the scattering parameter $\Gamma$ influences the electrodynamic response. 
Other theoretical approaches that also include a pair-breaking parameter but have a finite energy gap (though reduced compared to the BCS case without pair-breaking) could also explain some of our observations, e.g.\ they could model the difference between the experimental low-temperature conductivity spectra and the theoretical $\Gamma = 0$ curves in \autoref{fig:sigma-dynes}(a) for energies above $2\Delta$. Also, such different models, in the limit of small pair-breaking parameters as observed here, lead to very similar, small modifications to global superconducting parameters such as $T_\textrm{c}$, $n_\textrm{s,0}$, and $\Delta_0$ compared to the unperturbed case without pair breaking. 
But the non-vanishing below-gap $\sigma_1$ in the low-temperature limit, which features the observed step in $\sigma_1(f)$ at $E = \Delta$ in \autoref{fig:sigma-dynes}(a) and the specific temperature dependence in \autoref{fig:nu-tr-analysis}(a) for below-gap frequencies, requires sub-gap states near the Fermi energy, which are not consistent with only weak modifications of the density of states with only slightly reduced, finite energy gap.

Another possible explanation for sub-gap absorption in superconductors is collective modes, which have been the focus of previous THz studies on granular or strongly disordered superconductors \cite{Sherman2015, Pracht2017, Levy-Bertrand2019}. We think that such modes cannot explain our present observations, as their spectral signatures are very different \cite{Pracht2017, Levy-Bertrand2019} or they should become visible only for much stronger disorder than e.g.\ our \SI{20}{\nano\meter} film \cite{Sherman2015}.

\section{Summary}

We have measured the THz conductivity of five different superconducting NbN thin films with thicknesses ranging from $\SIrange{4.5}{20}{\nano\meter}$. 
For the thickest sample, we find in the real part $\sigma_1$ of the optical conductivity an absorption onset at $E \approx \Delta$, whose existence has been previously predicted for superconductors with Dynes DOS by Herman and Hlubina \cite{Herman2017}. 
Using both time- and frequency-domain THz techniques, we have validated and quantified this excess sub-gap conductivity.
Both measurements perfectly complement each other, yielding a temperature-independent pair-breaking rate of $\Gamma \approx 0.036 \Delta_0$. This shows the necessity of going beyond the Mattis-Bardeen and Zimmermann approaches to describe the optical response of this NbN thin film, and it confirms that the Dynes formalism can fully describe the data.
We find equivalent behavior also for thinner NbN films, see supplemental material,\cite{Supplemental} but there the $\sigma_1(f)$-step at $E \approx \Delta$ is less pronounced, in accordance with the experimentally determined smaller values of $\Gamma$ for the thinner films when compared to the \SI{20}{\nano\meter} sample. (Furthermore, experimental observation at $E \approx \Delta$ becomes more challenging for thinner films due to the relevant frequencies approaching the experimental TDS low-frequency limit and due to reduced overall THz absorption.) These results indicate that Dynes scattering is relevant in the THz regime at zero magnetic field, revealing a constant $\Gamma(T)$, in contrast to previous tunneling and point-contact measurements \cite{Chockalingam2009, Wu2020}.
This suggests that the microscopic processes that underlie the Dynes scattering parameter $\Gamma$ are more complex than previously thought. At the same time it demonstrates that the Dynes formalism might be the appropriate strategy to quantify sub-gap absorption that does not comply with Mattis-Bardeen predictions \cite{Herman2017, Simmendinger2016}, in particular for disordered superconductors such as transition-metal nitrides or carbides \cite{Baranek2025, Saritas2025} that are in the focus of both fundamental research (e.g.\ the superconductor-insulator transition) and applications (e.g.\ for superconducting quantum circuitry or cryogenic detectors) \cite{Oliver2013, EsmaeilZadeh2021, Wang2026}.

\begin{acknowledgments}
We thank M.\ Ziegler for supporting the sample growth, G.\ Untereiner for supporting the THz experiments, and C.\ Beydeda, F.\ Herman, and J.\ Pusskeiler for helpful discussions.
This work was supported by the Carl-Zeiss-Stiftung as part of CZS Center QPhoton and by the German Federal Ministry of Research, Technology and Space (BMFTR) within project QSolid,  Grant No.\ 13N16159. This work was partially supported by the BMFTR under Grant No.\ 13N16152/QSolid and ``NbNanoQ'', Grant No.\ 13N17121.
\end{acknowledgments}

\section*{Data availability}
The data that support the findings of this study are openly available from Zenodo \cite{ZenodoBolle}.

% Create the reference section using the .blg file that was previously generated from BibTeX:
%apsrev4-2.bst 2019-01-14 (MD) hand-edited version of apsrev4-1.bst
%Control: key (0)
%Control: author (8) initials jnrlst
%Control: editor formatted (1) identically to author
%Control: production of article title (0) allowed
%Control: page (0) single
%Control: year (1) truncated
%Control: production of eprint (0) enabled
%


\begin{thebibliography}{57}
\bibitem [{Sup()}]{Supplemental}%
  \BibitemOpen
  \href@noop {} {\bibinfo {title} {See {S}upplemental {M}aterial at [{URL} will
  be inserted by publisher] for additional {THz} data and analysis for all five
  samples as well as for a discussion of experimental error sources and how
  they affect the data analysis.}}\BibitemShut {Stop}%
\end{thebibliography}
\end{document}